\documentclass[sigconf]{acmart}

\AtBeginDocument{%
  \providecommand\BibTeX{{%
    \normalfont B\kern-0.5em{\scshape i\kern-0.25em b}\kern-0.8em\TeX}}}

\setcopyright{acmcopyright}
\copyrightyear{2018}
\acmYear{2018}
\acmDOI{10.1145/1122445.1122456}

\acmConference[Woodstock '18]{Woodstock '18: ACM Symposium on Neural
  Gaze Detection}{June 03--05, 2018}{Woodstock, NY}
\acmPrice{15.00}
\acmISBN{978-1-4503-XXXX-X/18/06}

\usepackage{color}
\usepackage{multirow}
\usepackage[para,online,flushleft]{threeparttable}

\usepackage{algorithm2e}
\usepackage{etoolbox}
\usepackage{xstring}
\usepackage{calc}
 
\newlength{\xalgowidth}
\newlength{\xalgoremainder}
\newlength{\xindentwidth}

\newenvironment{vAlgorithm*}[3][]{
  \setlength{\xalgowidth}{#2} 
  \setlength{\xindentwidth}{#3} 
  \setlength{\xalgoremainder}{\textwidth-\xalgowidth} 
  \SetCustomAlgoRuledWidth{\xalgowidth} 
  \IncMargin{\xindentwidth}
  \begin{algorithm*}[#1]
}
{
  \end{algorithm*} 
  \DecMargin{\xindentwidth}
}

{
  \end{algorithm} 
  \DecMargin{\xindentwidth}
}

\makeatletter
\patchcmd{\@algocf@start}{%
\begin{lrbox}{\algocf@algobox}%
}{%
\rule{0.5\xalgoremainder}{\z@}
\begin{lrbox}{\algocf@algobox}%
\begin{minipage}{\xalgowidth}%
}{}{}
\patchcmd{\@algocf@finish}{%
\end{lrbox}%
}{%
\end{minipage}%
\end{lrbox}%
}{}{}
\makeatother
\newcommand{\myalgorithm}{
	\LinesNumbered
	\DontPrintSemicolon
	\SetKwFunction{length}{length} \SetKwFunction{mymin}{min}
	\caption{AutoFAS: Automatic Feature and Architecture Selection for Pre-Ranking System}
    \label{alo:AutoFAS}
	\Indm
    \KwIn{$F$ and $R$ \quad set of input features and ranking network\newline
            $g$ and $M_i$ \quad feature mask and the $i^{\text{th}}$ Mixop\newline
            $L$, $S$, $T$ and $Lat$ \quad number of Mixops, initial step, total step and latency constraint}
	\Indp
	\BlankLine
	SelectedFeaturesAndArchitectures = $\emptyset$ \\
	\For{$i\leftarrow 1$ \KwTo $S$}{Train($F$, $R$) \quad \tcp{train a regular ranking model}} 
	SaveCheckpoint $R_0$ \quad \tcp{$R_0$ serves as teacher model} 
	\For{$i\leftarrow $S+1$ $ \KwTo $T$}{Train($g$, $M_{1:L}$; $R_0$, $F$, $Lat$) \quad \tcp{optimize the parameters in mask $g$ and $L$ by minimizing the \text{Loss} in Equ. \ref{eqn: losses}}}
	SelectedFeaturesAndArchitectures = Top($g$, $M_{1:L}$) \quad \tcp{select the largest strength features and architecture in each Mixop}
	$R_\text{pre-ranking}$ = Retrain(SelectedFeaturesAndArchitectures; $R_0$) \quad \tcp{distill the knowledge from $R_0$ to our searched pre-ranking model} 
	\KwOut{$R_\text{pre-ranking}$}
}

\begin{document}

\title{AutoFAS: Automatic Feature and Architecture Selection for Pre-Ranking System}

\author{Xiang Li}
\authornote{Both authors contributed equally to this research.}
\email{lixiang110@meituan.com}
\orcid{1234-5678-9012}
\author{Xiaojiang Zhou}
\authornotemark[1]
\email{zhouxiaojiang@meituan.com}
\affiliation{%
  \institution{Meituan Inc.}
  \streetaddress{4 East Wangjing Road}
  \city{Beijing}
  \country{China}
}

\author{Yao Xiao}
\email{xiaoyao06@meituan.com}
\affiliation{%
  \institution{Meituan Inc.}
  \streetaddress{4 East Wangjing Road}
  \city{Beijing}
  \country{China}}

\author{Peihao Huang}
\email{huangpeihao@meituan.com}
\affiliation{%
  \institution{Meituan Inc.}
  \streetaddress{4 East Wangjing Road}
  \city{Beijing}
  \country{China}
}

\author{Dayao Chen}
\email{chendayao@meituan.com}
\affiliation{%
  \institution{Meituan Inc.}
  \streetaddress{4 East Wangjing Road}
  \city{Beijing}
  \country{China}
}

\author{Sheng Chen}
\email{chensheng19@meituan.com}
\affiliation{%
  \institution{Meituan Inc.}
  \streetaddress{4 East Wangjing Road}
  \city{Beijing}
  \country{China}
}

\author{Yunsen Xian}
\email{xianyunsen@meituan.com}
\affiliation{%
  \institution{Meituan Inc.}
  \streetaddress{4 East Wangjing Road}
  \city{Beijing}
  \country{China}
}

\renewcommand{\shortauthors}{Li, et al.}

\begin{abstract}
Industrial search and recommendation systems mostly follow the classic multi-stage information retrieval paradigm: matching, pre-ranking, ranking, and re-ranking stages. To account for system efficiency, simple vector-product based models are commonly deployed in the pre-ranking stage. Recent works consider distilling the high knowledge of large ranking models to small pre-ranking models for better effectiveness. However, two major challenges in pre-ranking system still exist: (i) without explicitly modeling the performance gain versus computation cost, the predefined latency constraint in the pre-ranking stage inevitably leads to suboptimal solutions; (ii) transferring the ranking teacher's knowledge to a pre-ranking student with a predetermined handcrafted architecture still suffers from the loss of model performance. In this work, a novel framework AutoFAS is proposed which jointly optimizes the efficiency and effectiveness of the pre-ranking model: (i) AutoFAS for the first time simultaneously selects the most valuable features and network architectures using Neural Architecture Search (NAS) technique; (ii) equipped with ranking model guided reward during NAS procedure, AutoFAS can select the best pre-ranking architecture for a given ranking teacher without any computation overhead. Experimental results in our real world search system show AutoFAS consistently outperforms the previous state-of-the-art (SOTA) approaches at a lower computing cost. Notably, our model has been adopted in the pre-ranking module in the search system of Meituan \footnote{http://www.meituan.com}, bringing significant improvements.
\end{abstract}

\ccsdesc[500]{Information systems~Learning to rank}

\keywords{pre-ranking, feature and architecture selection, effectiveness, efficiency}



\maketitle

\section{Introduction}

\begin{figure}[h]
  \centering
  \includegraphics[width=0.7\linewidth]{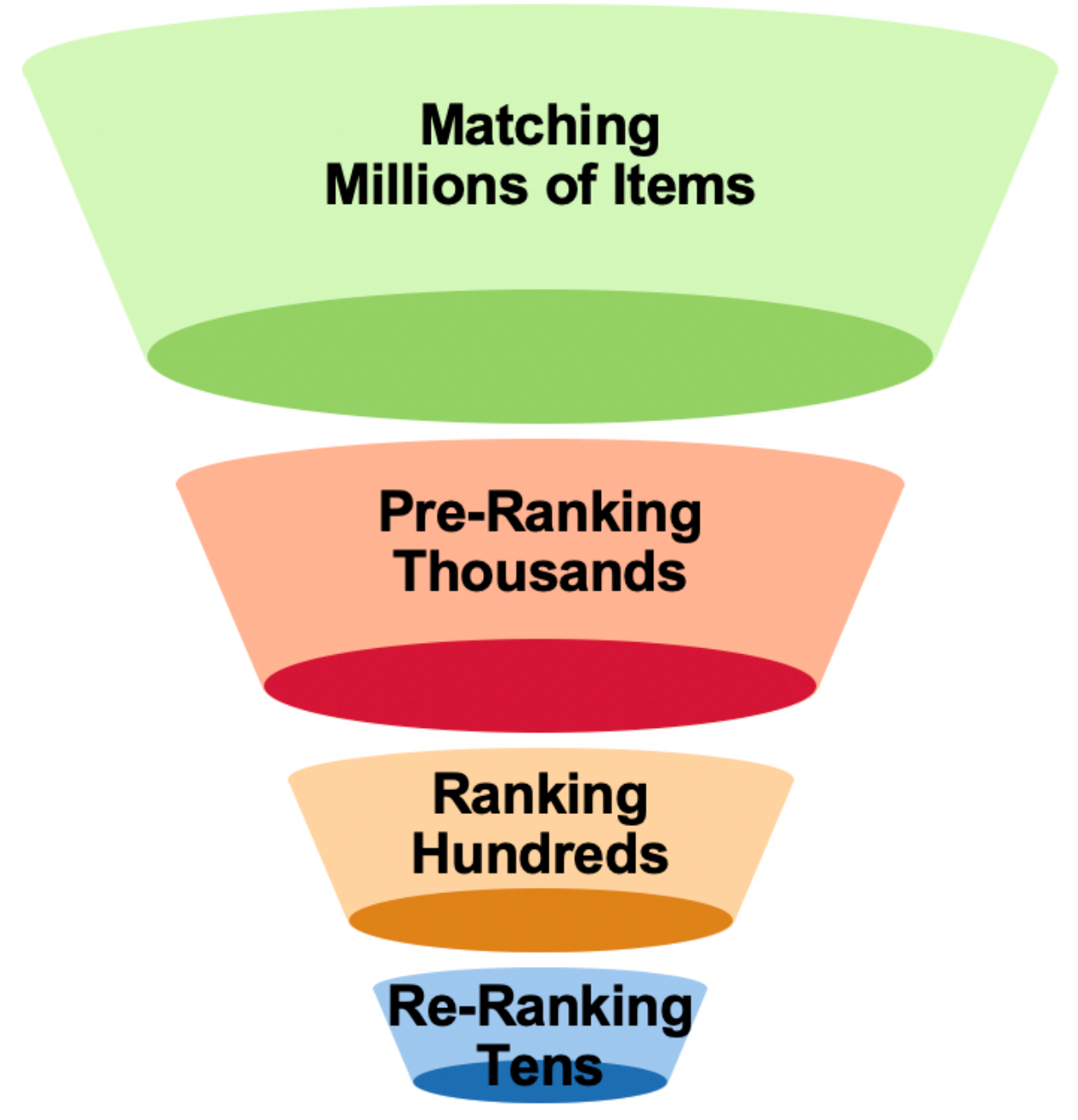}
  \caption{Real-world multi-stage ranking architecture with item numbers.}
  \label{fig:ranking architecture}
\end{figure}

Due to the information overload, search engine and recommendation system are becoming increasingly indispensable in assisting users to find their preferred items in web-scale applications such as Amazon and Meituan. As it is shown in the Fig.\ref{fig:ranking architecture}, a typical industrial searching system consists of four sequential stages: matching, pre-ranking, ranking and re-ranking. The effectiveness of search system not only influences the final revenue of whole platform, but also impacts user experience and satisfaction. In this paper, we mainly focus on the pre-ranking stage. 

There already exist numerous works on ranking models \cite{covington2016youtube, DBLP:journals/corr/ChengKHSCAACCIA16, zhou2018deep}. However, less attention is paid to pre-ranking models. The biggest obstacle to the development of pre-ranking system is the computation constraint. Taking the search engine of Meituan App for example. The size of the candidates to be scored for the pre-ranking system scales up to thousands, which is five times more than the subsequent ranking model. However, the latency limit is even more strict, e.g. 20 milliseconds at most. Specifically, approximately half of the latency is caused by feature retrieval, and the other half for model inference. Thus both features and architectures are needed to be optimized to achieve optimal results. 

To meet the computation constraint, logistic regression is a widely used lightweight model in the age of shallow machine learning. Due to the success of deep learning \cite{he2016deep}, representation-focused architectures \cite{10.1145/2959100.2959190} become dominant in the pre-ranking system. However, these representation-focused methods fail to utilize the interactive features, which turn out to be less efficient for computation but very effective for the expression ability \cite{wang2018billion}. Recently, COLD \cite{Wang2020COLDTT} firstly introduced interactive features into pre-ranking models. However, computation cost in COLD cannot be optimized jointly with model performance in an end-to-end manner. PFD \cite{10.1145/3394486.3403309} approaches this problem from a different angle by distilling the interactive features from more accurate teachers. But PFD does not take computation cost into account. Current state-of-the-art method FSCD \cite{10.1145/3404835.3462979} proposes a learnable feature selection method based on feature complexity and variational dropout. Nevertheless, as pointed in \cite{DBLP:journals/corr/abs-2007-00823}, variational dropout methods bias feature selection by ignoring hard-to-learn features. Moreover, none of these methods considers selecting the pre-ranking model architectures, which is also important for model efficiency. 

Inspired by the recent work \cite{pmlr-v80-pham18a, Cai2019ProxylessNASDN} in neural architecture search, we achieve a better trade-off between effectiveness and efficiency by designing a new pre-ranking methodology, which we name as AutoFAS: \textbf{Auto}matic \textbf{F}eature and \textbf{A}rchitecture \textbf{S}election for pre-ranking system. In terms of feature selection, we formulate it as a feature pruning process. Specifically, we first train a regular ranking model with all input features, mainly including user features, item features and interactive features. To automatically learn which feature is important, we explicitly introduce feature mask parameters to control whether its output should be passed to the next layer. Masked or not totally depends on the contribution of each feature to the final prediction. Those insignificant features are pruned out at the end of training for searched pre-ranking models. 

In parallel with feature selection, we relax the choices of candidate architectures to be continuous by introducing a set of architecture parameters (one for each architecture) so that the relative importance of each architecture can be learned by gradient descent. Similar to feature selection, only the top strength architectures are retained at the end of training for searched pre-ranking model. In order to distill the ranking teacher’s knowledge into both parameters and architecture of the pre-ranking student, we introduce a knowledge distillation(KD) loss during searching process. As will be shown in table \ref{tb2}, even trained on the same task and dataset, AutoFAS can select different optimal student architectures for different teachers and they consistently outperform conventional students with handcrafted architectures \cite{10.1145/3394486.3403309}. Finally, we model feature and architecture latency as a continuous function and optimize it as regularization loss with the aim of meeting the strict limitation of latency and computation resources. In our Table \ref{Effective performance of different pre-ranking model}, we show that compared to previous state-of-the-art results, AutoFAS is able to achieve 2.04\% improvement in AUC (Area Under Curve), 11\% improvement in Recall rate \cite{Wang2020COLDTT} and 1.22\% improvement in CTR (Click Through Rate) with a significant 10.3\% decrease in latency. A CTR lift of $0.1\%$ is considered significant improvement \cite{10.1145/3442381.3450078}.

To summarize, the main contribution of the paper can be highlighted as follows:

\begin{itemize}
    \item To the best of our knowledge, AutoFAS is the first algorithm that simultaneously learns features and architectures in search, recommendation and online advertising systems. In particular, it achieves a better trade-off between effectiveness and efficiency in pre-ranking stage.  
    \item AutoFAS successfully leverages Neural Architecture Search (NAS), equipped with our ranking model guided reward, to search for pre-ranking student that best aligned with subsequent ranking teacher. 
    \item Extensive experiments including online A/B test show the advantage of AutoFAS compared to previous state-of-the-art results. AutoFAS now has been successfully utilized in the main search engine of Meituan, contributing a remarkable business growth.
\end{itemize}


\section{Related Work}

\textbf{Pre-Ranking Methods} The structure of pre-ranking model has evolved from shallow to
deep. As a pioneer work of deep vector-based method, DSSM \cite{10.1145/2505515.2505665} trains a non-linear projection to map the query and the documents to a common semantic space, where the relevance is calculated as the cosine similarity between vectors. MNS \cite{50257} uses a mixture of batch and uniformly sampled negatives to tackle the selection bias in vector-based approaches. But as pointed in DIN \cite{10.1145/3219819.3219826}, the lack of interaction features between user and item significantly hampers the performance of vector-based pre-ranking models. With the increase of computing power, COLD \cite{Wang2020COLDTT} firstly introduces interactive features to pre-ranking system by pruning unimportant features. However, the trade-off between model performance and computation cost in COLD is decided offline, inevitably leading to inferior performance. FSCD \cite{10.1145/3404835.3462979} optimizes the efficiency and effectiveness in a learnable way. But it ignores the influence of underlying model structures. 

\noindent \textbf{NAS in Search and Recommendation System} Neural Architecture Search (NAS) has been an active research area since 2017 \cite{Zoph2017NeuralAS}. NIS \cite{10.1145/3394486.3403288} utilizes NAS to learn the optimal vocabulary sizes and embedding dimensions for categorical features. AutoFIS \cite{10.1145/3394486.3403314} formulates the problem of searching the effective feature interactions as a continuous searching problem using DARTS \cite{liu2018darts} technique. Via modularizing representative interactions as virtual building blocks and wiring them into a space of direct acyclic graphs, AutoCTR \cite{Song2020TowardsAN} searches the best CTR prediction model. AMEIR \cite{ijcai2021-290} focuses on automatic behavior modeling, interaction Exploration and multi-layer perceptron (MLP) Investigation. AutoIAS \cite{10.1145/3459637.3482234} unifies existing interaction-based CTR prediction model architectures and propose an integrated search space for a complete CTR prediction model. However, none of these works focuses on the pre-ranking models. 

\noindent \textbf{Knowledge Distillation in Search and Recommendation System} Ranking Distillation \cite{10.1145/3219819.3220021}  firstly adopts the idea of knowledge distillation to
large-scale ranking problems by generating additional training data and
labels from unlabeled data set for student model. Rocket Launching \cite{Zhou2018RocketLA} proposes a mutual learning style framework to train well-performing light CTR models. PFD \cite{10.1145/3394486.3403309} transfers the knowledge from a teacher model that additionally utilizes the privileged features to a regular student model. \cite{10.1145/3340531.3412704} explores the use of a powerful ensemble of teachers for more accurate CTR student model training. CTR-BERT \cite{Muhamed2021CTRBERTCK} present a lightweight cache-friendly factorized model for CTR prediction that consists of twin-structured BERT-like encoders using cross-architecture knowledge distillation. 


\begin{figure*}[t]
\centering
\includegraphics[width=0.8\linewidth]{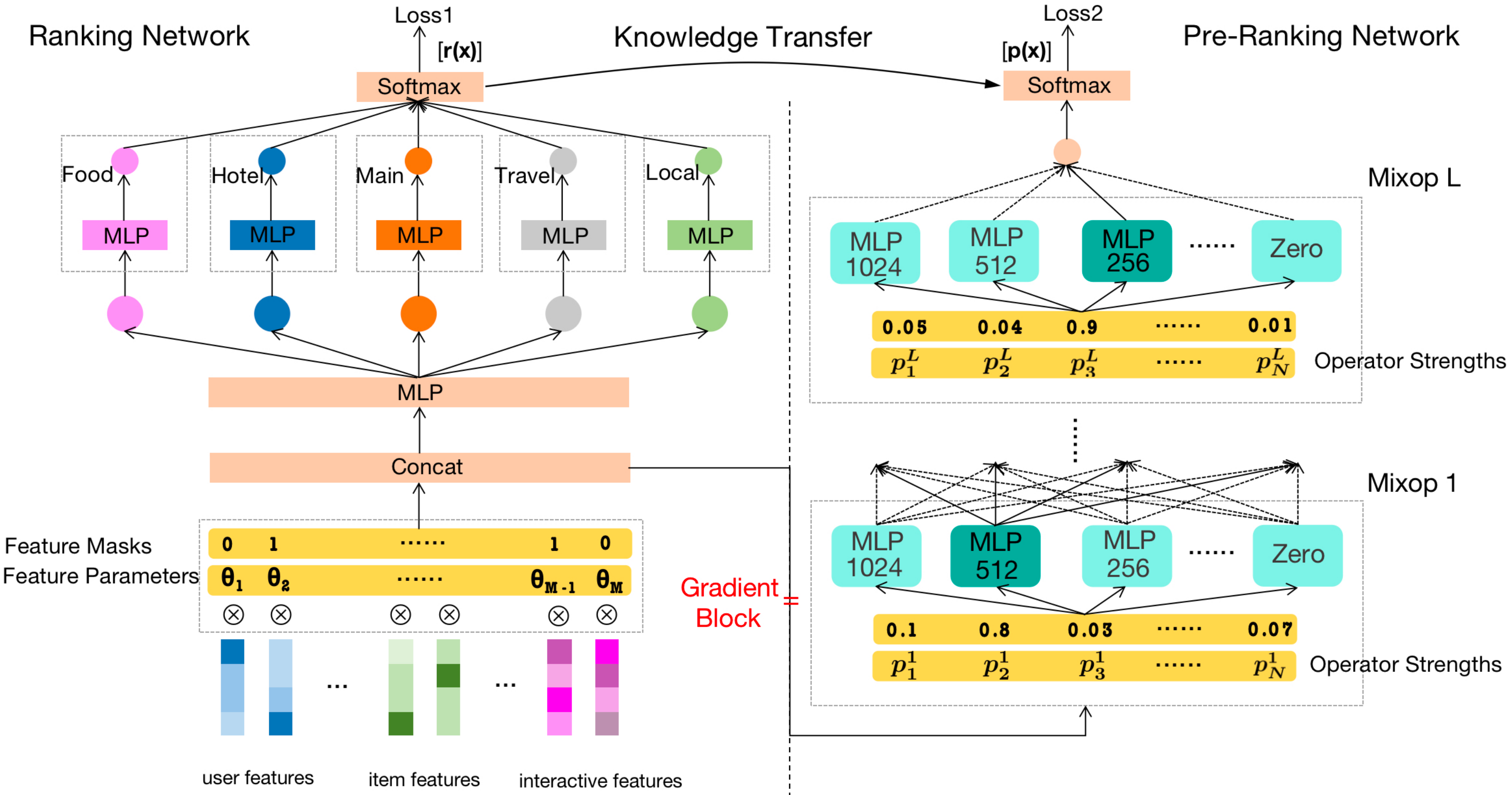} 
\caption{Network architecture of the proposed AutoFAS framework. AutoFAS is composed of two main parts. The left sub-network is our regular ranking network with feature mask module. Since the search engine of Meituan serves multiple business domains with overlapping user groups and items, our ranking model has multi-partition structure. The right sub-network consists of $L$ Mixops including all candidate pre-ranking architectures. The selected strongest operator in each Mixop which denoted in dark color forms the final architecture of the pre-ranking model.} 
\label{fig:nas_pic}
\end{figure*}

\section{Methods}
Our work is built on top of neural architecture search (NAS), thus we first present an overview of this topic. Then we will give a brief introduction of pre-ranking and describe our proposed methods for pre-ranking in detail. 

\subsection{Neural Architecture Search}
Neural network design requires extensive experiments by human experts. In recent years, there has been a growing interest in developing algorithmic NAS solutions to automate the manual process of architecture design \cite{Zoph2017NeuralAS, pmlr-v80-bender18a, Li2020ImprovingON}. Some works \cite{pmlr-v80-bender18a, pmlr-v80-pham18a} attempt to improve search efficiency via sharing weights across models, which further divided into two categories: continuous relaxation method \cite{liu2018darts, Cai2019ProxylessNASDN} and One-Shot method \cite{Brock2018SMASHOM, inbook}. Basically we follow the weight sharing methodology which includes three steps:(1) Design an over-parameterized network as search space containing every candidate architecture. (2) Make architectural decisions directly on the training set or a held-out validation set. (3) Re-train the most promising architectures from scratch and evaluate their performance on the test set. Notice that one big difference between our scenario and previous results is that we need to jointly search for both features and architectures at the same time. 

\subsection{Introduction of Search and Recommendation System}
The overall structure of search and recommendation system is already illustrated in Figure \ref{fig:ranking architecture}. Basically, the matching stage takes events from the user’s activity history as well as current query (if exists) as input and retrieves a small subset (thousands) of items from a large corpus (millions). These candidates are intended to be generally relevant to the user with moderate precision. Then the pre-ranking stage provides broad personalization and filters out top hundreds items with high precision and recall. Some companies may choose to combine matching and pre-ranking stages, like Youtube \cite{covington2016youtube}. Then the complex ranking network  assigns a score to each item according to a desired objective function using a rich set of features describing the item and user. The highest scoring items are presented to the user, ranked by their score, if without re-ranking. In general, pre-ranking shares similar functionality of ranking. The biggest difference lies in the scale of the problem. Directly applying ranking models in the pre-ranking system will face severe challenge of computing power cost. How to balance the model performance and the computing power is the core part of designing the pre-ranking system.

\subsection{Development History of Pre-Ranking in Meituan}
As mentioned before, pre-ranking module can be viewed as a transition stage between matching and ranking. Meituan is the largest Chinese shopping platform for locally found consumer products and retail services including entertainment, dining, delivery, travel and other services. At main search of Meituan, it receives thousands of candidates from matching stage and filters out top hundreds for the ranking stage. 
Our underlying pre-ranking architecture evolved from two-tower models \cite{10.1145/2505515.2505665}, Gradient Boosting Decision Tree (GBDT) models to the current deep neural network models during past years. As the performance increases, the excessive computational complexity and massive storage make it a greater challenge to deploy for real-time serving. The bottleneck of our online inference engine mainly contains two parts: feature retrieve from the database and deep neural network inference. Thus the feature selection and neural architecture selection are both important for the successful deployment of efficient and effective pre-ranking models. 

\subsection{Feature and Architecture Selection in Pre-Ranking}
One key motivation behind our approach is that we should co-build the pre-ranking model and subsequent ranking model such that the knowledge from ranking model can automatically guide us to find the most valuable features and architectures for pre-ranking model. Thus instead of training pre-ranking models separately, we co-train it with regular ranking model. We first describe the construction of the search space, then introduce how we leverage feature and architecture parameters to search for the most valuable features and architectures. Finally, we present our technique to handle the latency and KD-guided reward.

\textbf{Search Space} As shown in Fig. \ref{fig:nas_pic}, the left half of the graph is our ranking network, while the right half is the over-parametered network that contains all candidate pre-ranking models. The two parts share the same input features $F = \{ f_1, f_2, ..., f_M\}$. In our setup, $F$ mainly consists of user features, item features and interactive features. We train the standard ranking model with all $M$ feature inputs and then zero out large portions of features of the ranking model to evaluate their importance, choosing the best feature combinations. 

In parallel with feature selection, we need to search for the optimal architecture. Let $\mathcal{O}$ be a building block that contains $N$ different candidate operators: $\mathcal{O}=\{O_1, O_2, \cdots, O_N\}.$ In our case, $\mathcal{O}$ includes zero operator or multilayer perceptrons(MLP) \cite{Naumov2019DeepLR} with various hidden units.  Zero operator is the operator that keeps the input as output. Some references also consider it as identity operator. Notice that zero operator allows the reducing of number of layers. Other operators such as outer product \cite{inproceedingsdeepcross} and dot product could also be similarly abstracted and integrated into the framework, which is left for future exploration. To construct the over-parameterized network that includes every candidate architecture, instead of setting each edge to be a definite primitive operation, we set each edge to be a mixed operation (Mixop) that has N parallel paths, denoted as $m_O$. Then our over-parameterized network can be expressed as $\mathcal{N}(e_1 = m_O^1, \cdots, e_L = m_O^L)$, where $L$ is the total number of Mixops.

\textbf{Feature and Architecture Parameters} To select the most efficient features, We introduce $M$ real-valued mask parameters $\{\theta_i\}_{i=1}^M$, where $M$ is the number of features involved. Unlike \cite{10.5555/2969442.2969588} which binarizes individual weights, we binarize entire feature embedding. Thus the independent mask $g_i$ for feature $f_i$ is defined as the following Bernoulli distribution:
\begin{equation}
    g_i=\begin{cases}
			[1,\cdots, 1], & \text{with probability $\theta_i$}\\
            [0,\cdots, 0], & \text{with probability $1 - \theta_i$}
		 \end{cases}
\end{equation}
where the dimensions of $1$s and $0$s are determined by the embedding dimension of $f_i$. $M$ independent Bernoulli distribution results are sampled for each batch of examples. Since the binary masks $\{g_i\}_{i=1}^M$ are involved in the computation graph, feature parameters $\{\theta_i\}_{i=1}^M$ can be updated through backpropagation.

\begin{figure*}[t]
\centering
\includegraphics[width=0.8\linewidth]{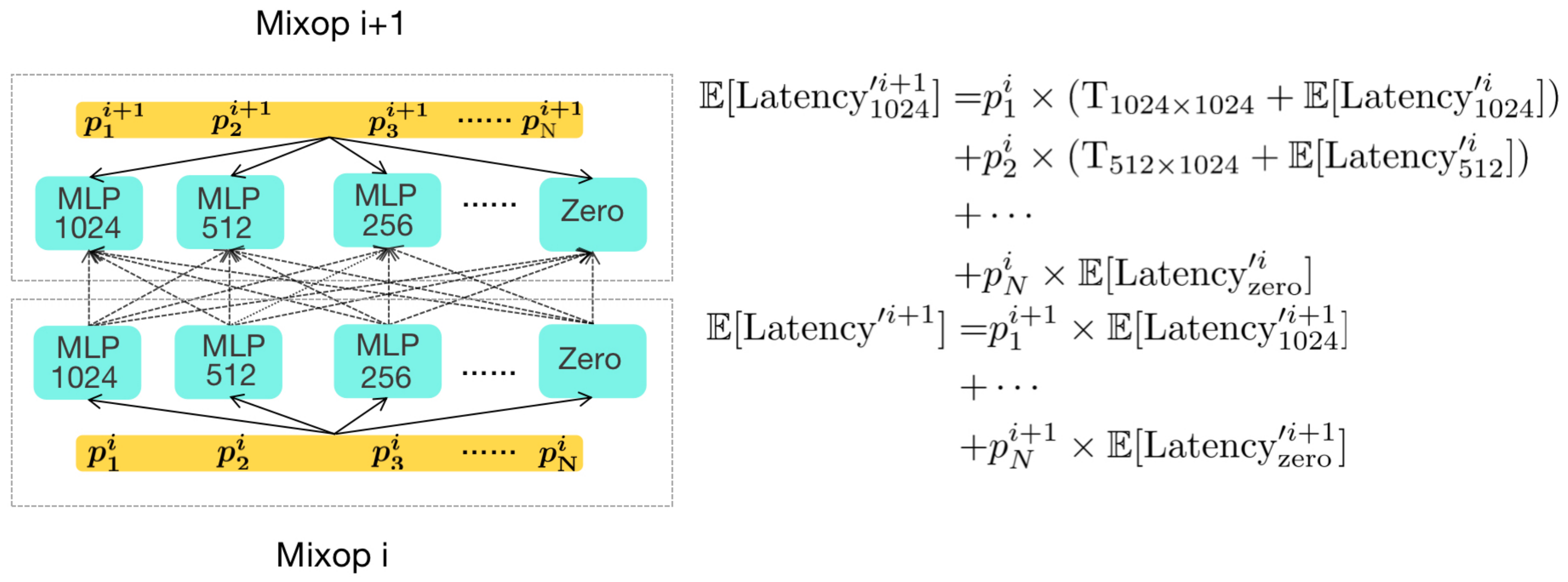} 
\caption{An example of computing the expected latency of each Mixop by recursion. Take the notation $T_{1024\times 1024}$ in above equation for illustration. It means the latency of a multi-layer perceptron with input dimension $1024$ and output dimension $1024$. It is counted by replaying the real-world request of our search engine to this particular network architecture. Every $p$ in the figure is the operator strength defined in equation \ref{eqn: lat}.} 
\label{fig: latency_pic}
\end{figure*}

In terms of architecture parameters, we will shown how to get the $N$ outputs of Mixop $i+1$, given the outputs of $N$ paths of Mixop $i$. As shown in Fig. \ref{fig: latency_pic}, denote the paths of Mixop $i$ as $m_O^i=\{O_1^i, O_2^i, \cdots,  O_N^i\}$, we introduce $N$ real-valued architecture parameters $\{\alpha_j^{i+1}\}_{j=1}^N$. Then the $k$-th output of Mixop $i+1$ is computed as follows:
\begin{equation}
\label{eqn: lat}
\begin{aligned}
O_k^{i+1} =& \sum_{j=1}^Np^{i+1}_j\text{MLP}_j^k(O_j^i) \\
=& \sum_{j=1}^N\frac{\exp{(\alpha_j^{i+1})}}{\sum_{m=1}^N\exp{(\alpha_m^{i+1})}}\text{MLP}_j^k(O_j^i)
\end{aligned}
\end{equation}
where the multi-layer perceptron $\text{MLP}^k$ has the same number of units as $O_k^{i+1}$,  $p^{i+1}_j := \frac{\exp{(\alpha_j^{i+1})}}{\sum_{m=1}^N\exp{(\alpha_m^{i+1})}}$ can be seen as the strength of the $j$-th operator in Mixop $i+1$. After this continuous relaxation, our goal is to jointly learn the architecture parameters and the weight parameters within all the mixed operations.

\textbf{Latency Constraint} 
Besides accuracy, latency (not FLOPs or embedding dimensions) is another very important objective when designing pre-ranking systems for real-world application. To make latency differentiable, we model the latency of a network as a continuous function of the neural network architectures. In our scenario, there exist two factors: feature related latency and architecture related latency. The features can be further divided into two categories from latency perspective: the ones passed from matching stage and the ones retrieved from in-memory dataset, denoted as $F_1$ and $F_2$ respectively. As such, we have the expected latency of a specific feature $f_i$:
\begin{equation}
    \mathbb{E}[\text{latency}_i] = \theta_i \times L_i
\end{equation}
where $L_i$ is the return time that can be recorded by the server. Then the gradient of $\mathbb{E}[\text{latency}_i]$ w.r.t. architecture parameters can thereby be given as: $\partial\mathbb{E}[\text{latency}_i] / \partial \theta_i = L_i$. Then the expected feature related latency of the network can be calculated as follows:
\begin{equation}
\begin{aligned}
\mathbb{E}[\text{latency}] =\max_{f_i \in F_1, f_j \in F_2} (&\mathbb{E}[\text{latency}_i] + \beta \cdot |F_1| ,  \\
&\mathbb{E}[\text{latency}_j] + \gamma \cdot|F_2|)
\end{aligned}
\end{equation}
where $|F_k|$ denotes the number of features in $F_k, k = 1, 2$, $\beta $ and $\gamma$ reflect the different concurrencies of the underlying system and can be decided empirically. 

We incorporate this expected feature latency into the regular loss function by multiplying a
scaling factor $\lambda$ which controls the trade-off between accuracy and latency. The final loss function for feature selection is given by: 
\begin{equation}
    \text{Loss1} = \text{Loss}_{\text{Ranking}}(y, f(X; \theta, W_{\text{Ranking}})) + \lambda \mathbb{E}[\text{latency}]
\end{equation}
where $f$ denotes the ranking network.

Similarly, for architecture latency of Mixop $i+1$, we can compute its expected latency $\mathbb{E}[\text{latency}'^{i+1}]$ by recursion, as shown in right figure of Fig. \ref{fig: latency_pic}. Since these operations are executed sequentially during inference, the expected latency of the pre-ranking network can be expressed as the expected latency of the last Mixop:
\begin{equation}
    \mathbb{E}[\text{latency}'] = \mathbb{E}[\text{latency}'^{\text{L}}]
\end{equation}

\textbf{Ranking System Supervision} Knowledge distillation \cite{inproceedings}, the process of transferring the generalization ability of the teacher model to the student, has recently received increasing attention from the research community and industry. 
While conventional one-hot label in supervision learning constrains the 0/1 label, the soft probability output from teacher model contributes to the knowledge for student model. 
Remember that one drawback of current KD method \cite{10.1145/3394486.3403309} in pre-ranking system is that it only transfers the teacher's knowledge to a student with fixed neural architecture. Inspired by the success of AKD \cite{inproceedingsLiu}, we propose to add a distillation loss to the architecture search process. Specifically, we employ the soft targets produced by ranking models as the supervision signal to guide the selection in each Mixop. Thus the final loss function for architecture selection is given:
\begin{equation}
\begin{aligned}
\text{Loss2} = & (1-\lambda_1) \text{Loss}_{\text{pre-Ranking}}(y, g(X; \theta, \alpha, W_{\text{pre-Ranking}})) \\
& + \lambda_1 ||r(x) - p(x)|| ^ 2_2 + \lambda_2 \mathbb{E}[\text{latency}']
\end{aligned}
\end{equation}
where $g$ is the pre-ranking network, $\text{Loss}_{\text{pre-Ranking}}$ denotes the pre-ranking pure loss with the known hard labels $y$. $r(x)$ and $p(x)$ are final softmax activation outputs of ranking and pre-ranking network, respectively. 
\RestyleAlgo{algoruled}
\begin{vAlgorithm*}[ht!]{\textwidth}{0.5in}
\myalgorithm
\end{vAlgorithm*}

We will further discuss the effectiveness of $\lambda_1$ and distillation loss in section \ref{sec: as_analysis}. $\lambda_2$ is the scaling factor that controls the trade-off between accuracy and latency. Loss1 and Loss2 are optimized together, resulting in the final multi-task loss function:

\begin{equation}
\label{eqn: losses}
 \text{Loss} = \text{Loss1} + \text{Loss2}
\end{equation}

The absence of balancing hyperparameter between Loss1 and Loss2 comes from that Loss1 only optimizes feature mask parameters, while Loss2 optimizes architecture parameters and weights in the pre-ranking model. We choose this strategy because it is empirically better than the model without gradient block (both feature parameters and architecture parameters can be optimized by Loss2), shown in Table \ref{tab: ablation}. Loss1 and Loss2 are related to each other by the fact that the input of Loss2 is the masked embedding, where the mask parameters are continuously optimized by Loss1 during training. To derive the final pre-ranking architecture, we retain the strongest features and operators in each Mixop and retrain it from scratch. The whole training process of AutoFAS can be summerized in Algorithm \ref{alo:AutoFAS}.

\section{Experiments}
\label{sec: experiments}
In this section, we present the experiment results in detail. First, we introduce experiment datasets, training details and evaluation metric. Then we compare our proposed AutoFAS with competitors in terms of both feature selection and architecture selection. Finally, we discuss the effectiveness of critical technical designs in AutoFAS through ablation study. 

\subsection{Datasets}
To the best of our knowledge, there is no public dataset for pre-ranking task. Previous works \cite{Wang2020COLDTT, 10.1145/3404835.3462979} in this area present results in their own dataset. To verify the effectiveness of AutoFAS, we conduct experiments on the industrial dataset of Meituan. It is collected from the platform searching system of Mobile Meituan App. Samples are constructed from impression logs, with 'click' or 'not' as label. This dataset contains more than 10 billion display/click logs of 20 million users and 400 million 'click' in 9 days. To alleviate the sample selection bias in pre-ranking, we preprocess these impression samples by adding non-displayed examples, depending on the sample orders in later ranking model. Training set is composed of samples from the first 7 days, validation and test set are from the following 2 days, a classic setting for industrial modeling. 

\subsection{Experimental Settings}
We choose the size $M$ of feature set as $500$, mainly including user features, item features and interactive features. To build architecture space, we allow $L = 5$ Mixops, including multi-layer perceptron (MLP) with various units $\{1024, 512, 256, 128, 64\}$. To enable a direct trade-off between width and depth, we add zero operation to the candidate set of its mixed operation. In this way, with a limited latency budget, the network can either choose to be shallower and wider by skipping more blocks and using MLPs with more unit or choose to be deeper and thinner by keeping more blocks and using MLPs with less units. The size of joint search space for both feature and architecture can be approximated as $2^{500} \times 6^5 \approx 10^{155}$. 

We first train our ranking network for $S = 6$ million steps without any mask to obtain reasonable embedding weights for input features. Then we continue regular optimization of Loss with respect to mask parameters $\theta$, architecture parameters $\alpha$ and weights in pre-ranking networks. The optimizer is Adagrad with learning rate 0.01 and the batch size is 50. Note that the feature embedding parameters and weights in ranking networks are fixed after initial 6 million steps. 

The training cost of a pre-ranking model like COLD \cite{Wang2020COLDTT} in our setup is 2 days, while AutoFAS needs to be trained for 3 days from scratch. In practice, the training cost could be reduced to 2 days by loading a well-trained ranking model, which is the common case in industry. Thus our AutoFAS methods will not add much training overhead compared to previous SOTA methods. 

In terms of effectiveness measurement, we use three popular indexes: AUC (Area Under Curve) and Recall \cite{Wang2020COLDTT} as offline metric, CTR (Click Through Rate) as online metric. Notice that the Recall serves to measure the alignment degree between the pre-ranking model and subsequent ranking model. 
For evaluation of system performance, we use metrics including RT (return time, which measures the latency of model) and CPU consumption rate metrics. All models reported in this paper run on a CPU machine with Intel(R) Xeon(R) CPU E5-2650 v4 @ 2.20GHz (12 cores) with 256GB RAM. Generally speaking, lower RT and CPU consumption means lower computation cost. 

\subsection{Baselines}
\begin{itemize}
    \item \textbf{VPDM} \cite{50257}: VPDM (Vector-Product based DNN Model) is a widely used method at the early stage of deep CTR prediction task. It maps the query and candidate items into common low-dimensional space where the ranking score of a item is readily computed as the inner product between the query and this item.
    \item \textbf{COLD} \cite{Wang2020COLDTT}: Besides some engineered optimization tricks on its specific platform, COLD calculates the importance of every feature by Squeeze-and-Excitation block \cite{10.1109/TPAMI.2019.2913372} and selects the top ones based on the computation cost. 
    \item \textbf{FSCD} \cite{10.1145/3404835.3462979}: FSCD is the current state-of-the-art pre-ranking architecture that learns efficient features by considering feature complexity and variational dropout. 
\end{itemize}

\begin{table}
\caption{Performance of different feature selection methods for pre-ranking system. Note that the latency in the table refers to the feature related latency.}
\label{tb1}
\centering
\setlength{\tabcolsep}{1.1mm}{
\begin{tabular}{c|c|c|c|c|c}
\hline                      
\multicolumn{1}{c|}{\bf N}  &\multicolumn{1}{c|}{\bf Method} &\multicolumn{1}{c|}{\bf AUC} &\multicolumn{1}{c|}{\bf Recall} &\multicolumn{1}{c|}{\bf CPU}&\multicolumn{1}{c}{\bf Latency}\\ 
\hline
  &COLD   &0.7495 &0.33   &22\% &6.32 ms\\
50    &FSCD  &0.7503 &0.35   &21\% &5.75 ms\\
    &AutoFAS & \bf0.7505 &\bf0.36 &\bf17\% &\bf4.59 ms\\
\hline
  &COLD   &0.8001  &0.52  &30\% &7.27 ms\\
100    &FSCD  &0.8009  &0.52  &29\% &7.19 ms\\
    &AutoFAS &\bf0.8012 &\bf0.53 &\bf26\% &\bf5.87 ms\\
\hline
120 &AutoFAS &0.8270 &0.62 &30\% &7.28 ms\\
\hline
  &COLD   &0.8274 &0.63  &39\% &9.18 ms\\
150    &FSCD  &0.8283 &0.65  &39\% &9.11 ms\\
    &AutoFAS &\bf0.8285 &\bf0.67 &\bf35\% &\bf7.99 ms\\
\hline
\end{tabular}
}
\end{table}

\begin{table}[t!]
\caption{Performance comparison of top-100 features selected by AutoFAS and Statistics\_AUC}
\label{Performance comparision between features choosen by our model}
\setlength{\tabcolsep}{0.5mm} {
\begin{tabular}{c|c|c|c|c|c}
\hline                      
\multicolumn{1}{c|}{\bf N} &\multicolumn{1}{c|}{\bf Method}  &\multicolumn{1}{c|}{\bf AUC} &\multicolumn{1}{c|}{\bf Recall} &\multicolumn{1}{c|}{\bf CPU} &\multicolumn{1}{c}{\bf Latency} \\ 
\hline
\multirow{2}{*}{100} &Statistics\_AUC &0.7781 &0.50 &28\% &7.03 ms\\
 &AutoFAS &\bf0.8012 &\bf0.53 &\bf26\% &\bf5.87 ms\\
\hline
\end{tabular}
}
\end{table}

\subsection{Analysis of Feature Selection}
\textbf{Overall Result} To fairly compare the effects of our feature selection algorithm with others, we fix a 3-layers’ MLP, with the number of hidden neurons being 512 and 256, as the common pre-ranking structure for all methods. Table \ref{tb1} lists the model effectiveness and system efficiency for all models with different number N of features. 
Particularly, when ${N=100}$, AutoFAS beats the current state-of-the-art result FSCD by 0.03\% and 1.0\% in terms of AUC and Recall, respectively. However, the latency is 18.4\% lower than FSCD. If we relax our model to have approximately the same latency and CPU consumption rate as FSCD, we can keep top-120 features and obtain significant performance boost by 2.61\% and 10.0\% in terms of AUC and Recall, respectively. We also observes that, when $N > 120$, the AUC and Recall increase slowly, while the latency and CPU cost increase remarkably.

\noindent \textbf{Detailed Examination} In this part, we will investigate the effectiveness of features selected by different methods. Inspired by AutoFIS \cite{10.1145/3394486.3403314}, we use \textbf{statistics\_AUC} to represent the contribution of one single feature to the final prediction. For a given feature, we evaluate a well trained predictor with all feature inputs except for this one. Then the decrease of AUC is referred to as statistics\_AUC of this feature. Thus higher statistics\_AUC indicates a greater impact on the final prediction.

\begin{figure}[t]
\centering
\includegraphics[width=1.0\linewidth]{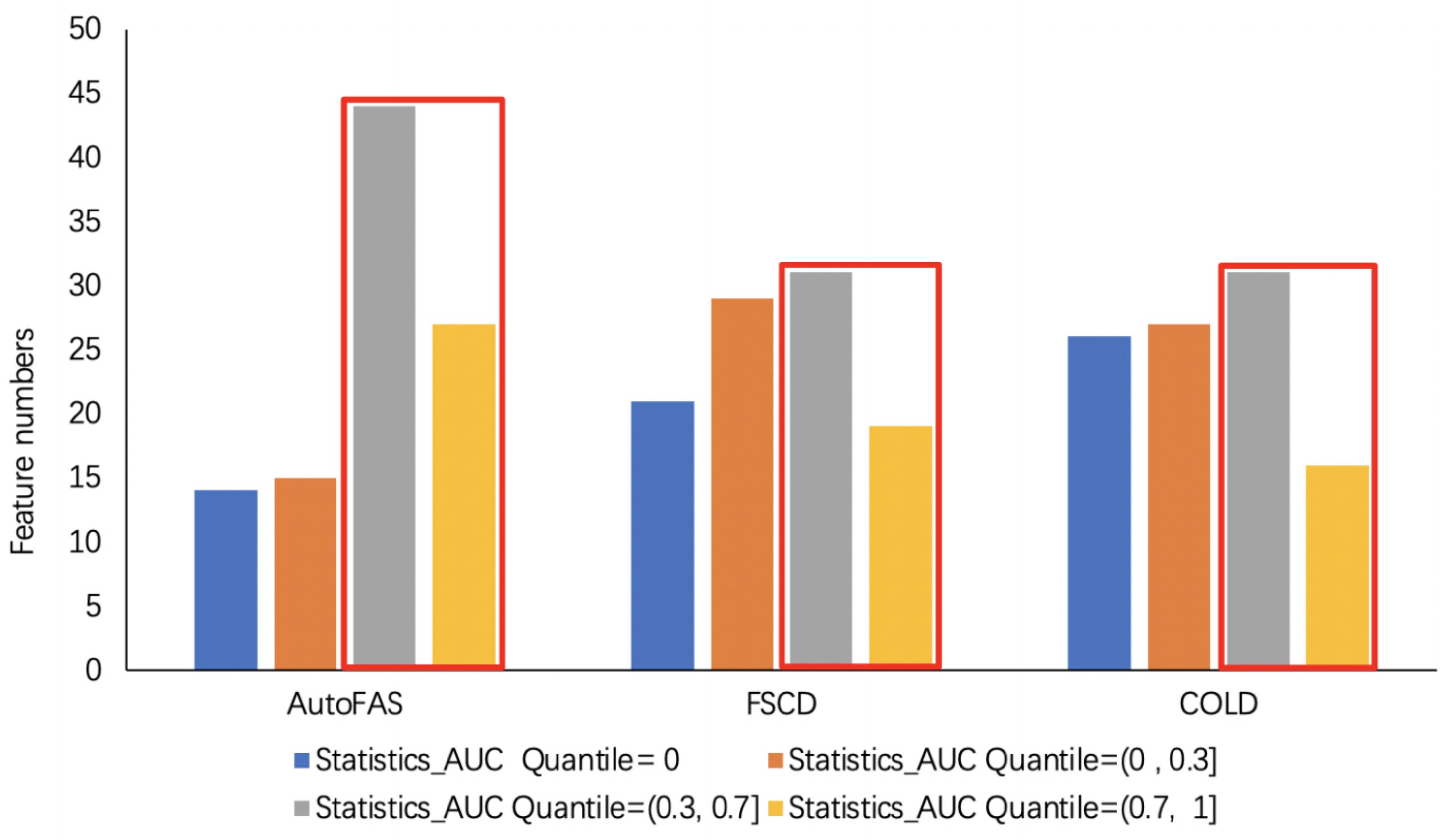}
\caption{Statistics\_AUC distribution of different methods}
\label{fig: feature_valid}
\end{figure}

\begin{table*}
\caption{Performance comparison of handcrafted pre-ranking architectures with different students searched by different teachers.}
\label{tb2}
\centering
  \begin{tabular}{c|c|c|c|c}
    \hline
    \multirow{2}{*}{Teachers} &
      \multicolumn{3}{c|}{AUC} &
      \multirow{2}{*}{Comparison} \\
      \cline{2-4}
    & Student1 & Student2 & handcrafted \\
    \hline
    Teacher1($\lambda_1=0.2$) & \bf 0.8576  & 0.8559 & 0.8551 & handcrafted \textless student2 \textless \  \bf student1\\
    \hline
    Teacher2($\lambda_1=0.6$) & 0.8457 & \bf 0.8471 & 0.8447  & handcrafted \textless student1 \textless \  \bf student2\\
    \hline
  \end{tabular}
\end{table*}

We can visualize the statistics\_AUC distribution of different methods in Figure \ref{fig: feature_valid}. Notice that the quantile is referred to the top-100 statistics\_AUC features and 0 quantile means not among top-100. As it is shown, AutoFAS can select more high statistics\_AUC features than FSCD and COLD. Specially, we find that some high statistics\_AUC features like userItemDis (the distance between user and item) and userGeoItemView (the history interaction between user and item) are among top-100 of AutoFAS, but not FSCD or COLD. Since Meituan is a e-commerce platform for local services, the LBS (location based service) features like userItemDis and userGeoItemView are definitely of importance. A natural follow-up question to ask is that how about keeping all the top-100 features based on statistics\_AUC. We list the results in Table \ref{Performance comparision between features choosen by our model}. We argue that there exist too much duplicated information in top statistics\_AUC features, leading to inferior performance conjointly.

\subsection{Analysis of Architecture Selection}
\label{sec: as_analysis}
In this part, we investigate the power of our architecture selection module. The experiment is designed as follows. In Table \ref{tb2} we have two teachers, which are the same ranking model $R_0$ with different distillation hyperparameter $\lambda_1$. Besides the teacher model, the other settings, random seed, latency target, input features and dataset are fixed to be the same. As is shown in Table \ref{tb2}, student1 and student2 are two student architectures searched by two teachers, respectively. We also add the popular handcrafted architecture which has decreasing width (the detailed architectures of these three models are shown in Figure \ref{fig: structure}). We observe that while student1 outperforms student2 with teacher1 as teacher, student2 works better with teacher2 as teacher, implying different teachers could have different best students. Note that the different best students supervised by different teachers consistently outperform the handcrafted architecture. This result indicates that our architecture selection module in the knowledge distillation is indeed necessary and can further boost the effectiveness without any overhead. We choose student1 as our final architecture. 

\begin{table}[t]
\caption{Performance of different pre-ranking models. Notice that the CTR is reported through a online A/B test and latency is the entire latency including feature retrieval latency and model inference latency.}
\label{Effective performance of different pre-ranking model}
\setlength{\tabcolsep}{1mm} {
\begin{tabular}{c|c|c|c|c|c}
\hline                      
\multicolumn{1}{c|}{\bf Method} &\multicolumn{1}{c|}{\bf AUC} &\multicolumn{1}{c|}{\bf Recall} &\multicolumn{1}{c|}{\bf CTR} &\multicolumn{1}{c|}{\bf CPU} &\multicolumn{1}{c}{\bf Latency}    \\ 
\hline
VPDM   &0.7535  &0.49 & - &\bf46\% &\bf18.7ms\\
COLD   &0.8350  &0.70 &+0.34\%     &51\% &22.8ms\\
FSCD   &0.8372  &0.72 &+0.58\%    &51\%   &22.4ms\\
AutoFAS  &\bf0.8576  &\bf0.83 &\bf+1.80\%   &48\% &20.1ms\\
\hline
\end{tabular}
}
\end{table}

\subsection{Analysis of Feature and Architecture Selection}
\begin{figure}[t]
\centering
\includegraphics[width=0.6\linewidth]{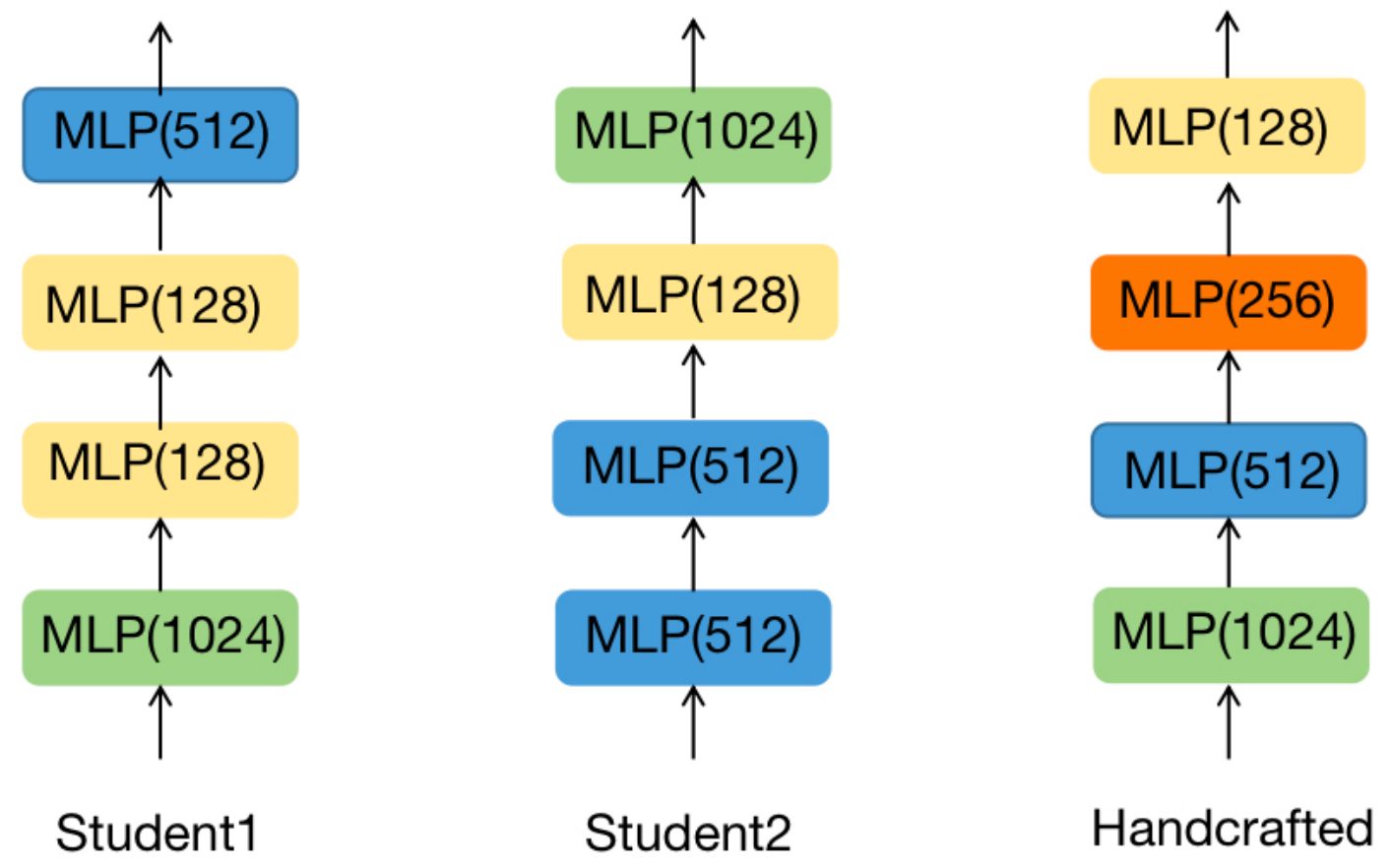}
\caption{Selected and handcrafted neural architectures.}
\label{fig: structure}
\end{figure}

In this experiment, we compare AutoFAS with all baseline models including VPDM, COLD and FSCD (all with 100 features). The dimension in VPDM is chosen to be 32. The underlying architecture for COLD and FSCD is the handcrafted one in Figure \ref{fig: structure}. Apart from offline results, we also conduct a strict online A/B testing experiment to validate the proposed AutoFAS model, from 2021-07-18 to 2021-07-24. 10\% of the total traffic is distributed for each model. As is shown in Table \ref{Effective performance of different pre-ranking model}, AutoFAS achieves great gain of 1.8\% in CTR in Meituan main search scene. In terms of system efficiency, compare to current state-of-the-art model FSCD, AutoFAS decrease the latency by 10.3\% with a surprising absolute CTR lift of 1.22\%, which is significant to the business.

\subsection{Ablation Study}
\label{subsec: ablation}

\subsubsection{Sensitivity of Hyper-parameter} In the above experiments, we first train our ranking network without any mask for $S=6$M global steps. Here we test the sensitivity of $S$. The result is shown in Table \ref{tab: ablation}. Different global steps can make a noticeable difference. Static global steps 6M improve the model performance than 3M significantly. However, 10M has only slightly performance improvement than 6M at the cost of almost double training time.

\subsubsection{Effects of Training Manner} In the above experiments, we utilize the gradient block technique to cancel the effect of distillation loss’s back-propagation on feature mask parameters. Moreover, to maintain the maximal knowledge, the teacher $R_0$'s output is produced without feature masks. Here we test the effects of such training manners. The result is also shown in Table \ref{tab: ablation}. We can see discarding the gradient block technique brings 0.68\% decrease in AUC. If we acquire the teacher output with feature masks, the AUC degenerates notably 0.95\%. These two results together imply that the current training manner lead us much better performance  while adding no burden to training time.






\section{Conclusions}
In this paper, we device an end-to-end AutoML pre-ranking pipeline AutoFAS. Instead of simply considering feature combinations, AutoFAS simultaneously selects both features and model architectures. The joint optimization of computation cost and model performance ensures a better trade-off between effectiveness and efficiency. Furthermore, our tailored neural architecture search algorithm with KD-guided reward empowers AutoFAS knowledge from subsequent cumbersome ranking models. Experimental results on the real-world dataset demonstrate the effectiveness of the proposed pre-ranking approach. Future work includes enriching search space, exploring more efficient search strategies as well as automatic distributing computation power among matching, pre-ranking, ranking and re-ranking modules.

\begin{table}[t]
\caption{Comparisons of different framework design’s result.}
\label{tab: ablation}
\setlength{\tabcolsep}{0.9mm} {
\begin{threeparttable}
\begin{tabular}{c|c|c|c|c}
\hline                      
\multicolumn{1}{c|}{\bf Method} &\multicolumn{1}{c|}{\bf AUC} &\multicolumn{1}{c|}{\bf Recall} &\multicolumn{1}{c|}{\bf CPU} &\multicolumn{1}{c}{\bf Latency}    \\ 
\hline
AutoFAS(S=6M)\tnote{1} &0.8576 &0.83  &48\% &20.1ms\\
AutoFAS(S=3M) &0.8552 &0.82   &47\% &19.4ms\\
AutoFAS(S=10M) &0.8579 &0.83   &48\% &19.8ms\\
\hline
AutoFAS(w/o GB)\tnote{2}  &0.8508  &0.80  &48\% &20.3ms\\
\hline
AutoFAS(w FM)\tnote{3} &0.8481 &0.78   &48\% & 19.6ms\\
\hline
\end{tabular}
\begin{tablenotes}
\item[1] base model \\ 
\item[2] w/o GB means removing gradient block technique \\ 
\item[3] w FM means teacher inference with feature masks
\end{tablenotes}
\end{threeparttable}
}
\end{table}

\bibliographystyle{ACM-Reference-Format}
\bibliography{sample-base}










\end{document}